\documentclass[aps,preprint]{revtex4}%
\usepackage{amsfonts}
\usepackage{amsmath}
\usepackage{amssymb}
\usepackage{graphicx}%
\providecommand{\U}[1]{\protect\rule{.1in}{.1in}}

\begin{document}
\preprint{ }
\title[1/N ]{Soft-photon exponentiation beyond the quenched approximation in QED$_{2+1}$ }
\author{Yuichi Hoshino}
\affiliation{Kushiro National College of Technology,Otanoshike Nishi
2-32-1,Kushiro,Hokkaido, 084-0916,Japan}

\begin{abstract}
We discuss the infrared and ultraviolet behavior of the fermion propagator in
(2+1)-dimensional QED based on spectral representation.If we choose
soft-photon exponentiation to include all orders of soft-photon emission by
electron,its spectral function may be written as $e^{F}$,where $F$ is a model
independent spectral function of the lowest order in the coupling constant.We
evaluate the function $F$ in an analytic way and show its short and long
distance behavior with an infrared cut-off $\mu$.At short distance function
$F$ has linear and logarithmic infrared divergence.However in the long
distance limit function $F$ vanishes.So that only short distance part of
$e^{F}$ is modified from unity.We may avoid the linear divergence by the
choice of the gauge $d=-1$,where $d$ is a covariant gauge fixing parameter.In
this gauge the spectral function vanishes in the limit of zero bare photon
mass $\mu$.We overcome this difficulty by adding continuous spectrum of
massive fermion loop to photon spectral function $\rho(\mu^{2})$,where $\mu$
has the role of invariant mass for fermion-antifermion pair and is larger than
$2m.$So that unquenched fermion spectral function survibes.For the application
of chiral symmetry breaking we carefully studied the position space propagator
$S_{F}(0)$.At least for weak coupling these values agree quite well with that
obtained in Dyson-Schwinger equation with proper vertex correction.We also
study these parameter as a function of the flavour number $N$ and t'Hooft
coupling $\alpha=e^{2}N/8\pi$ for strong coupling case.

\end{abstract}
\volumeyear{year}
\volumenumber{number}
\issuenumber{number}
\eid{identifier}
\date[Date text]{date}
\received[Received text]{date}

\revised[Revised text]{date}

\accepted[Accepted text]{date}

\published[Published text]{date}

\startpage{101}
\endpage{102}
\maketitle
\tableofcontents

\section{ Introduction}

About 30 years ago it was pointed out that high temperature limit of the field
theory is described by the same theory with less-dimension and it suffers from
severe infrared divergences with dimensionful coupling constant [1,2]. From
this results we may feel that the asymptotic field does not exist for
QED$_{3}$ by severe infrared singularities.In this work we discuss the
infrared behaviour of the massive fermion propagator by so called soft photon
exponentiation.For the S matrix and the transition probability soft-photon
exponentiation is familiar by the work of S.Weinberg.It has been famous to
prove the cancellation of infrared divergences between real photon process and
virtual photon process in the transition probability in QED[3].There is a
renormalization group analysis as well as Bloch-Nordsieck model on the
infrared behaviour of the electron propagator.These works are known to lead
one particle singularity of the propagator,where infrared anomalous dimension
modifies the pole structure of electron[8].It has been shown that approximate
propagator by soft photon exponentiation leads the same infrared bahavior as
the renormalization group analysis[3].Following soft-photon exponentiation we
discuss the femion propagator in (2+1)-dimensional QED.In (2+1)-dimensional
QED, the leading infrared divergence for $O(e^{2})$ spectral function
$F(x,\mu)$ is\ linear as $1/\mu,$where $\mu$ is a bare photon mass.This term
turned out to be proportional to $(d+1)$,we may choose gauge parameter $d=-1$
to avoid linear infrared divergences.These soft photon contributions may be
expressed as an extra factor as $e^{F(x,\mu)}.$In this way we obtain the
fermion position space full propagator by $S_{F}^{0}(x)e^{F(x,\mu)}$.From this
form of the propagator we show only short distance part of the propagator is
modified while the long distance part is kept as free one by soft-photon
correction with finite infrared cut-off $\mu$.This is the main feature in our
approximate solution to the propagator in the previous work[4].In this case we
have a serious problem : the vanishment of the propagator in the limit $\mu$
equals to zero.In this limit $e^{F}$ vanishes as $(\mu|x|)^{\gamma}$ for small
$|x|$ ,as $\exp(-\gamma m/\mu\exp(-\mu|x|))$ for large $|x|$ where
$\gamma=e^{2}/8\pi m$.These fact show a severe infrared confinement in
quenched case.To obtain finite propagator with zero infrared cut-off we may
replace bare photon propagator with full propagator including the vacuum
polarization of massive fermion loops.In this case imaginary part of photon
propagator provides photon spectral function $\rho(\mu^{2})$ with $2m\leq
\mu\leq\infty$.In section II we introduce specral representation of fermion
and photon. Perturbative and non-perturbative spectral functions based on its
definition are given too.Section III is devoted to the analysis in position
space propagator.We evaluate the full propagator for quenched case with bare
photon mass and improve it in unquenched case.In section IV using full
propagator we show the vacuum expectation value $\left\langle \overline{\psi
}\psi\right\rangle $ as a function of coupling constant and numbers of flavor
$N_{f}$,renormalization constant $Z_{2}$,and the results obtained by
Dyson-Schwinger equation with vertex correction which satisfy
Ward-Takahashi-identity.Section V is devoted to Summary.

\section{ \ Spectral representation of the propagator}

\subsection{\bigskip Fermion}

In this section we show how to evalute the fermion propagator non
perturbatively by the spectral represntation which preservs unitarity
analyticity and CPT invariance [2,3,4,7].Assuming parity conservation we adopt
4-component spinor.The spectral function of the fermion in (2+1) dimension is
defined
\begin{align}
\left\langle 0|T(\psi(x)\overline{\psi}(y)|0\right\rangle  &  =i\int
\frac{d^{3}p}{(2\pi)^{3}}e^{-ip\cdot(x-y)}\int_{0}^{\infty}ds\frac{\gamma\cdot
p\rho_{1}(s)+\rho_{2}(s)}{p^{2}-s+i\epsilon},\\
\rho(p)  &  =\frac{1}{\pi}\operatorname{Im}iS_{F}(p)=\gamma\cdot p\rho
_{1}(p)+\rho_{2}(p)\nonumber\\
&  =(2\pi)^{2}\sum_{n}\delta^{(3)}(p-p_{n})\left\langle 0|\psi
(0)|n\right\rangle \left\langle n|\overline{\psi}(0)|0\right\rangle .
\end{align}
In the quenched approximation the state $|n>$ stands for a fermion and
arbitrary numbers of photons,%
\begin{equation}
|n>=|r;k_{1},...,k_{n}>,r^{2}=m^{2}.
\end{equation}
In deriving the matrix element $\left\langle 0|\psi(0)|n\right\rangle $ we
must take into occount the soft photon emission vertex which is written in the
textbook for the scattering of charged particle by external electromagnetic
fied or collision of charged particles.Based on low-energy theorem the most
singular contribution for the matrix element $T_{n}=\left\langle \Omega
|\psi|r;k_{1},....,k_{n}\right\rangle $ is known as the soft photons attached
to external line.We want to consider $T_{n}$ for $k_{n}^{2}\neq0,$hence we
continue off the photon mass shell by a Lehmann-Symanzik-Zimmermann(LSZ)
formula.First we notice the Fourier expansion of $A_{in}(x)$ as that of free
fields.%
\[
\mathbf{A}_{in}^{T}(x)=\int d^{2}k[a_{in}^{T}(k)\mathbf{A}_{k}^{T}%
(x)+a_{in}^{+T}(k)\mathbf{A}_{k}^{\ast T}(x)
\]
with%
\begin{equation}
\mathbf{A}_{k}^{T}(x)=\frac{1}{(2\pi)^{2}2k_{0}}e^{-ik\cdot x}\mathbf{\epsilon
}^{T}(k)
\end{equation}
and, upon inversion,%
\begin{align}
a_{in}(k,\lambda)  &  =i\int d^{2}x\mathbf{A}_{k}^{\ast T}(x)\cdot
\overleftrightarrow{\partial_{0}}\mathbf{A}_{in}^{T}(x)\nonumber\\
&  =-i\int d^{2}xA_{k,\lambda}^{\ast}(x)_{\mu}\cdot\overleftrightarrow
{\partial_{0}}A_{in}(x)^{\mu}%
\end{align}
where $\overleftrightarrow{\partial_{0}}$ is defined by%
\begin{equation}
f\overleftrightarrow{\partial_{0}}g=f\partial_{0}g-(\partial_{0}f)g.
\end{equation}
In developing the reduction formula for photon ,there is a minor change from
transverse photon to arbitrary state%
\begin{align}
&  \frac{1}{\sqrt{Z_{3}}}\mathbf{A}_{k_{i}}^{T}(x_{i})\overrightarrow{\square
}_{x_{i}}\cdot\left\langle 0|\cdot\cdot\cdot\mathbf{A}^{T}(x_{i})\cdot
\cdot\cdot|0\right\rangle \nonumber\\
&  =-\frac{1}{\sqrt{Z_{3}}}A_{k_{i},\lambda_{i}}^{\mu}(x_{i})\overrightarrow
{\square_{x_{i}}}\left\langle 0|\cdot\cdot\cdot A_{\mu}(x_{i})\cdot\cdot
\cdot|0\right\rangle .
\end{align}
The additional minus sign in (5),(6) comes from the space-like nature of the
polaization unit vector%
\[
\epsilon_{\mu}\epsilon^{\mu}=-\mathbf{\epsilon}\cdot\mathbf{\epsilon}=-1.
\]%
\begin{align}
\epsilon_{\mu}^{n}T_{n}^{\mu}  &  =\left\langle \Omega|T\psi a_{\mu}%
^{+n}(k)|r;k_{1},k_{2},....,k_{n-1}\right\rangle \nonumber\\
&  =-i\lim_{t_{i}\rightarrow-\infty}\int_{t_{i}}d^{2}xe^{ik_{n}\cdot
x}\overleftrightarrow{\partial_{0}}\left\langle \Omega|T\psi\epsilon_{\mu}%
^{n}A_{\mu}^{in}(x)|r;k_{1},k_{2},..,k_{n-1}\right\rangle \nonumber\\
&  =-i\int d^{3}x\partial_{0}[e^{ik_{n}\cdot x}\overleftrightarrow
{\partial_{0}}\left\langle \Omega|T\psi\epsilon_{\mu}^{n}A_{\mu}%
^{in}(x)|r;k_{1},k_{2},....k_{n-1}\right\rangle ].
\end{align}
Noting that $\partial_{0}(f\overleftrightarrow{\partial_{0}}%
g)=f\overleftrightarrow{\partial_{0}^{2}}g$ and $\ \partial_{0}^{2}%
e^{ik_{n}\cdot x}=\nabla^{2}e^{ik_{n}\cdot x}$,we find%
\begin{align}
\epsilon_{\mu}^{n}T_{n}^{\mu}  &  =\frac{-i}{\sqrt{Z_{3}}}\int d^{3}%
xe^{ik_{n}\cdot x}\square_{x}\left\langle \Omega||T\psi\epsilon_{\mu}%
^{n}A_{\mu}^{n}(x)|r;k_{1},k_{2},...,k_{n-1}\right\rangle \nonumber\\
&  =\frac{-i}{\sqrt{Z_{3}}}\int d^{3}xe^{ik_{n}\cdot x}\left\langle
\Omega|T(\psi\epsilon_{\mu}^{n}j_{\mu}(x)+(1-d)\epsilon_{\mu}^{n}\partial
_{\mu}\partial\cdot A(x))|r;k_{1},k_{2},...,k_{n-1}\right\rangle .
\end{align}
provided
\begin{equation}
\square_{x}T\psi A_{\mu}(x)=T\psi\square_{x}A_{\mu}(x)=T\psi(j_{\mu
}(x)+(1-d)\partial_{\mu}^{x}\partial\cdot A(x)),
\end{equation}
where the electromagnetic current is
\begin{equation}
j_{\mu}(x)=e\overline{\psi}(x)\gamma_{\mu}\psi(x)
\end{equation}
and $d$ is a covariant gauge fixing parameter.From the definition (9) $T_{n}$
is seen to satisfy the following Ward-Takahashi identity
\begin{equation}
k_{\mu}^{n}T_{n}^{\mu}(r;k_{1},k_{2},...k_{n})=eT_{n-1}(r;k_{1},k_{2}%
,...k_{n-1}),r^{2}=m^{2},
\end{equation}
provided by current conservation
\begin{align}
\partial_{\mu}j_{\mu}(x)  &  =0,\nonumber\\
ik_{n}e^{ik_{n}\cdot x}T(\psi j_{\mu}(x))  &  =\partial_{x}^{\mu}%
(e^{ik_{n}\cdot x}T(\psi j_{\mu}(x)))-ie^{ik_{n}\cdot x}\partial_{x}^{\mu
}T(\psi j_{\mu}(x)),
\end{align}
equal time commutation relations
\begin{align}
\partial_{\mu}^{x}T(\psi j_{\mu}(x))  &  =-e\psi(x),\nonumber\\
\partial_{\mu}^{x}T(\overline{\psi}j_{\mu}(x))  &  =e\overline{\psi}(x),
\end{align}
and
\begin{equation}
\square\partial\cdot A(x)=0.
\end{equation}
\qquad\qquad Here we impose on the physical state%
\begin{equation}
\partial_{\mu}A^{\mu(+)}(x)|phys>=0,
\end{equation}
to drop the gauge dependent part.We have an approximate solution of equation
(12)
\begin{equation}
T_{n}|_{k_{n}^{2}=0}=e\frac{\gamma}{\gamma\cdot(r+k_{n})-m}T_{n-1}.
\end{equation}
From this relation the n-photon matrix element is replaced by the products of
$T_{1}$
\begin{equation}
T_{n}T_{n}^{+}\gamma_{0}\rightarrow%
{\displaystyle\prod\limits_{j=1}^{n}}
T_{1}(k_{j})T_{1}^{+}(k_{j})\gamma_{0}/n!.
\end{equation}
by one-photon state matrix element
\begin{align}
T_{1}  &  =\left\langle 0|\psi(0)|r,k\right\rangle =-i\left\langle
0^{in}|T[\psi(0),e\int d^{3}x\overline{\psi}^{in}(x)\gamma_{\mu}\psi
^{in}(x)A_{\mu}^{in}(x)]|r;k\text{ in}\right\rangle \nonumber\\
&  =-ie\int d^{3}xS_{F}(0-x)\gamma_{\mu}\left\langle 0|\psi(x)|r\right\rangle
\left\langle 0|A_{\mu}(x)|k\right\rangle \nonumber\\
&  =\frac{-ie}{\gamma\cdot(r+k)-m+i\epsilon}\gamma_{\mu}\epsilon_{\lambda
}^{\mu}(k)U_{S}(r)\sqrt{\frac{m}{E_{r}}}\frac{1}{\sqrt{2k_{0}}},
\end{align}
where $U_{S}(r)$ is a free particle spinor[5].Spectral function $\rho$ is
written symbolically as%
\begin{align}
\rho(p^{2})  &  =\int\frac{d^{3}xe^{-ip\cdot x}}{(2\pi)^{3}}\int
(m+i\gamma\cdot\partial_{r})d^{3}r\delta(r^{2}-m^{2})e^{ir\cdot x}\nonumber\\
&  \times\sum_{n=0}^{\infty}(\int\frac{d^{3}k}{(2\pi)^{2}}e^{ik\cdot x}%
\delta(k^{2})\theta(k_{0}))\sum_{\lambda})_{n}T_{n}\overline{T}_{n},
\end{align}
where we replaced the $\delta$ function
\begin{equation}
\delta^{(3)}(p-r-k)\rightarrow\int\frac{d^{3}x}{(2\pi)^{3}}e^{-i(p-r-k)\cdot
x}.
\end{equation}
The polarization sum for photon is given
\begin{equation}
\sum_{\lambda}\epsilon_{\lambda}^{\mu}(k)\epsilon_{\lambda}^{\nu}%
(k)=-[g^{\mu\nu}+(d-1)\frac{k^{\mu}k^{\nu}}{k^{2}}].
\end{equation}
Since we do not fix the numbers of photon $n$ we sum up from $n\ $equals zero
to infinity.The function
\begin{equation}
F=\int\frac{d^{3}k}{(2\pi)^{2}}\theta(k_{0})\delta(k^{2})2k_{0}\frac{E_{r}}%
{m}\sum\limits_{\lambda,S}T_{1}\overline{T_{1}}e^{-ik\cdot x},
\end{equation}
is a phase space integral of one photon intermediate state which may be
exponentiated in (20) as $e^{F}$.The soft photon extra factor is well known in
the evaluation of S matrix with soft photon emission[3].Accordingly spectral
function $\rho$ becoms
\begin{equation}
\rho(p^{2})=\int\frac{d^{3}xe^{-ip\cdot x}}{(2\pi)^{3}}\int\frac
{(m+i\gamma\cdot\partial_{r})d^{2}r}{2E_{r}}(e^{ir\cdot x}-e^{-ir\cdot
x})e^{F},
\end{equation}
which may be explained in the next section.

\subsection{Evaluation of the spectral function}

\subsubsection{\bigskip general property of spectral function}

Here we consider the fermion spectral function.The vacuum expectation value of
the anticommutator has the form[7]%

\begin{align}
S^{\prime}(x,y)  &  =i\left\langle 0|\{\psi(x),\overline{\psi}%
(y)\}|0\right\rangle \nonumber\\
&  =i\sum_{n}[\left\langle 0|\psi(0)|n\right\rangle \left\langle
n|\overline{\psi}(0)|0\right\rangle e^{-ip_{n}\cdot(x-y)}+\left\langle
0|\overline{\psi}(0)|n\right\rangle \left\langle n|\psi(0)|0\right\rangle
e^{ip_{n}\cdot(x-y)}].
\end{align}
We introduce the spectral amplitude by grouping togather in the sum over $n$
all states of given three-momentum $q$%
\begin{equation}
\rho_{\alpha\beta}(q)=(2\pi)^{2}\sum_{n}\delta^{(3)}(p_{n}-q)\left\langle
0|\psi_{\alpha}(0)|n\right\rangle \left\langle n|\overline{\psi}_{\beta
}(0)|0\right\rangle
\end{equation}
and set out to construct its general form from invariance arguments.$\rho(q)$
is a $4\times4$ matrix and may be expanded in terms of $16$ linearly
independent products of $\gamma$ matrices.Under the assumptions of Lorentz
invariance and Parity transformation it reduces to the form%
\begin{equation}
\rho(q)_{\alpha\beta}=\rho_{1}(q)\gamma\cdot q+\rho_{2}(q)\delta_{\alpha\beta
}.
\end{equation}
Second term in (25) can be related directly to (26) with the aid of PCT
invariance of the theory[7].For definitness we define $\gamma$-matrices by
Dirac representation:%
\begin{equation}
\gamma^{0}=\left(
\begin{array}
[c]{cc}%
I & 0\\
0 & -I
\end{array}
\right)  ,\gamma^{i}=\left(
\begin{array}
[c]{cc}%
0 & \sigma^{i}\\
-\sigma^{i} & 0
\end{array}
\right)  ,\gamma^{5}=\left(
\begin{array}
[c]{cc}%
0 & I\\
I & 0
\end{array}
\right)  ,\{\gamma^{\mu},\gamma^{\nu}\}=2g^{\mu\nu}%
\end{equation}
where $i=1,2,3,I$ is $2\times2$ identity matrix and the $\sigma^{i\prime}$s
are the Pauli matrices%
\begin{equation}
\sigma^{1}=\left(
\begin{array}
[c]{cc}%
0 & 1\\
1 & 0
\end{array}
\right)  ,\sigma^{2}=\left(
\begin{array}
[c]{cc}%
0 & -i\\
i & 0
\end{array}
\right)  ,\sigma^{3}=\left(
\begin{array}
[c]{cc}%
1 & 0\\
0 & -1
\end{array}
\right)  .
\end{equation}
Parity,Charge conjugation and Time\ reversal transformation are defined
\begin{align}
P\psi(t,x,y)P^{-1}  &  =\gamma^{1}\gamma^{5}\psi(t,-x,y),\\
PA^{0}(t,x,y)P^{-1}  &  =A^{0}(t,-x,y),\\
PA^{1}(t,x,y)P^{-1}  &  =-A^{1}(t,-x,y),\\
PA^{2}(t,x,y)P^{-1}  &  =A^{2}(t,-x,y),
\end{align}%
\begin{align}
C\psi(x)C^{-1}  &  =C\gamma^{0}\psi^{\ast}=C\overline{\psi}^{T}=\psi_{c},\\
C  &  =i\gamma^{2}\gamma^{0},C^{-1}\gamma^{\mu}C=-\gamma^{\mu T},\\
CA^{\mu}(t,x,y)C^{-1}  &  =-A^{\mu}(t,x,y),
\end{align}%
\begin{align}
T\psi(t,x,y)T^{-1}  &  =i\gamma^{1}\psi(-t,x,y),T=i\gamma^{1},\\
TA^{0}(t,x,y)T^{-1}  &  =A^{0}(-t,x,y),\\
TA^{1,2}(t,x,y)T^{-1}  &  =-A^{1,2}(-t,x,y).
\end{align}
Effects of PCT transformation on $\overline{\psi}(y)\psi(x)$ is%
\begin{align}
PCT\psi_{\alpha}^{A}(t,x,y)T^{-1}C^{-1}P^{-1}  &  =(\gamma^{5}\gamma^{2}%
\gamma^{0})_{\alpha\gamma}\overline{\psi}_{\gamma}^{T}(-t,-x,y),\\
PCT\overline{\psi}_{\beta}^{B}(t,x,y)T^{-1}C^{-1}P^{-1}  &  =(\gamma^{5}%
\gamma^{2}\gamma^{0})_{\lambda\beta}\psi_{\lambda}^{T}(-t,-x,y),\\
PCT\overline{\psi}_{\alpha}^{A}(t^{\prime},x^{\prime},y^{\prime})\psi_{\beta
}^{B}(t,x,y)T^{-1}C^{-1}P^{-1}  &  =(\gamma^{0}\gamma^{2}\gamma^{5}%
)_{\alpha\gamma}\psi_{\gamma}^{B}(-t,-x,y)\overline{\psi}_{\lambda}%
^{A}(-t^{\prime},-x^{\prime},y^{\prime})(\gamma^{0}\gamma^{2}\gamma
^{5})_{\lambda\beta}.
\end{align}
Inserting (42) along with (27) into (25 ) and using $\gamma^{2T}=-\gamma
^{2},\gamma^{5T}=\gamma^{5}$, we obtain finally%

\begin{align}
S_{\alpha\beta}^{\prime}(x-y)  &  =i\int\frac{d^{3}q}{(2\pi)^{3}}\theta
(q_{0})([\gamma\cdot q\rho_{1}(q^{2})+\rho_{2}(q^{2})]_{\alpha\beta
}e^{-iq\cdot(x-y)}\nonumber\\
&  +\{\gamma^{0}\gamma^{2}\gamma^{5}[\gamma\cdot q\rho_{1}(q^{2})+\rho
_{2}(q^{2})]\gamma^{0}\gamma^{2}\gamma^{5}\}_{\alpha\beta}e^{iq\cdot
(x^{\prime}-y^{\prime})})\nonumber\\
&  =i\int\frac{d^{3}q}{(2\pi)^{2}}\theta(q_{0})[\rho_{1}(q^{2})i\gamma
\cdot\partial_{x}+\rho_{2}(q^{2})]_{\alpha\beta}(e^{-iq\cdot(x-y)}%
-e^{iq\cdot(x^{\prime}-y^{\prime})})\nonumber\\
&  =i\int\frac{d^{3}q}{(2\pi)^{2}}[\theta(q_{0})\gamma\cdot q\rho_{1}%
(q^{2})+\epsilon(q_{0})\rho_{2}(q^{2})]_{\alpha\beta}e^{-iq\cdot(x-y)},
\end{align}
where $x^{\prime}=(-t,-x_{1},x_{2}).$Since $\rho$ vanishes for space-like
$q^{2}$,we may also write this as an integral over mass spectrum by
introducing%
\begin{equation}
\rho(q^{2})=\int_{0}^{\infty}\rho(s)\delta(q^{2}-s)ds.
\end{equation}
We find
\begin{align}
S^{\prime}(x-y)  &  =-\int ds[i\rho_{1}(s)\gamma\cdot\partial+\rho
_{2}(s)]\Delta(x-y;\sqrt{s})\nonumber\\
&  =\int ds\{\rho_{1}(s)S^{\prime}(x-y;\sqrt{s})+[\sqrt{s}\rho_{1}(s)-\rho
_{2}(s)]\Delta(x-y;\sqrt{s})\}
\end{align}
where invariant $\Delta$ function is given%
\begin{align}
\Delta^{\prime}(x,y)  &  \equiv-i\left\langle 0|[\phi(x),\phi
(y)]|0\right\rangle \\
&  =-i\sum_{n}\left\langle 0|\phi(0)|n\right\rangle \left\langle
n|\phi(0)|0\right\rangle (e^{-iP_{n}\cdot(x-y)}-e^{iP_{n}\cdot(x-y)}%
)\nonumber\\
&  =\frac{-i}{(2\pi)^{2}}\int d^{3}q\rho(q^{2})\theta(q_{0})(e^{-iq\cdot
(x-y)}-e^{iq\cdot(x-y)})\\
&  =\frac{-i}{(2\pi)^{2}}\int_{0}^{\infty}ds\rho(s)\int d^{3}q\delta
(q^{2}-s)\epsilon(q_{0})e^{-iq\cdot(x-y)}\nonumber\\
&  =\int_{0}^{\infty}ds\rho(s)\Delta(x-y,\sqrt{s}).
\end{align}
From the\ above representation of propagator $S^{\prime}(x-y)$ we have the
relation $\rho_{2}(s)=\sqrt{s}\rho_{1}(s)$.The above spectral representation
goes through unchanged for the vacuum expectation value of the time-ordered
product of Dirac field;it is necessary only to replace the $S$ and $\Delta$ by
the Feynman propagator $S_{F}$ and $\Delta_{F}$.

\subsubsection{\bigskip non-perturbative spectral function}

Here we return to our case.In our approximation,propagator is a product of
free one and the exponentiation of one photon state $e^{F}$.First we set
$\rho_{1}(r^{2})=\delta(r^{2}-m^{2}),\rho_{2}(r^{2})=\sqrt{r^{2}}\delta
(r^{2}-m^{2}).$Second using on-shell momentum expansion,its Fourier
transformation is given
\begin{equation}
\rho(p^{2})=\int\frac{d^{3}xe^{-ip\cdot x}}{(2\pi)^{3}}\int\frac
{d^{2}r(m+i\gamma\cdot\partial)}{2E_{r}}(e^{-ir\cdot x}-e^{ir\cdot x}%
)e^{F(x)}.
\end{equation}
We evaluate the above in the $p$ rest frame,by writing the integral in the
form%
\begin{align}
\rho(p^{2})  &  =\int_{-\infty}^{\infty}\frac{dte^{-ip_{0}t}}{2\pi}\int
d^{2}x\int\frac{d^{2}r(m+i\gamma_{0}\cdot\partial_{t})}{(2\pi)^{2}2E_{r}%
}(e^{ir\cdot\mathbf{x}-iE_{r}t}-e^{-ir\cdot\mathbf{x}+iE_{r}t})e^{F(x)}%
\nonumber\\
&  =\int_{-\infty}^{\infty}\frac{dte^{-ip_{0}t}}{2\pi}\int d^{2}x\int
\frac{d^{2}r}{(2\pi)^{2}}(\frac{m}{2E_{r}}(e^{ir\cdot\mathbf{x-}iE_{r}%
t}-e^{-ir\cdot\mathbf{x+}iE_{r}t})\nonumber\\
&  +\frac{\gamma_{0}E_{r}}{2E_{r}}(e^{ir\cdot\mathbf{x}-iE_{r}t}%
+e^{-ir\cdot\mathbf{x}+iE_{r}t}))e^{F(x)}.
\end{align}
We expand%
\[
\frac{1}{2E_{r}}e^{iE_{r}t}e^{F(x)}%
\]
in powers of $r$ keeping only the first terms in $\rho$ as%
\[
\frac{1}{2E_{r}}e^{iE_{r}t}e^{F(x)}=\frac{e^{imt}}{2m}e^{F(it)}.
\]
Performing $r$ integral gives%
\begin{align}
\rho_{1}(p^{2})  &  =\int_{-\infty}^{\infty}\frac{dt}{2\pi}\frac{1}%
{2m}(e^{i(m-p_{0})t}+e^{-i(m+p_{0})t})\int d^{2}x\delta^{2}(x)e^{F(it)}%
,\nonumber\\
&  =\int_{-\infty}^{\infty}\frac{dt}{2\pi}\frac{1}{2m}(e^{i(m-p_{0}%
)t}+e^{-i(m+p_{0})t})e^{F(it)},\\
\rho_{2}(p^{2})  &  =\int_{-\infty}^{\infty}\frac{dt}{2\pi}\frac{|p_{0}|}%
{2m}(e^{i(m-p_{0})t}+e^{-i(m+p_{0})t})e^{F(it)}.
\end{align}
$\rho_{1}(p^{2}),\rho_{2}(p^{2})$ are mormalized to $\delta(p^{2}-m^{2})$ and
$\sqrt{p^{2}}\delta(p^{2}-m^{2})$ when $F(t)=0.$In this way we obtain%
\begin{align}
\rho_{1}(s)  &  =\int_{-\infty}^{\infty}\frac{dt}{2\pi}e^{-i(s-1)m^{2}%
t}e^{F(it)},s=p^{2}/m^{2},\\
\rho_{2}(s)  &  =\int_{-\infty}^{\infty}\frac{dt}{2\pi}m\sqrt{s}%
e^{-i(s-1)m^{2}t}e^{F(it)}.
\end{align}
Now we evaluate the spectral function.First we examine the four dimensional
scalar case as a guide. It has been shown in [3] that the function $F$ in
four-dimension was derived in the same way in three dimension as
\begin{equation}
D^{+}(x)=\frac{1}{(2\pi)^{3}i}\int e^{ik\cdot x}\theta(k^{0})\delta
(k^{2}-m^{2})d^{4}k.
\end{equation}%
\begin{align}
D^{+}(x)  &  =\frac{1}{4\pi}\epsilon(x^{0})\delta(x^{2})-\frac{mi}{8\pi
\sqrt{x^{2}}}\theta(x^{2})[N_{1}(m\sqrt{x^{2}})-i\epsilon(x^{0})J_{1}%
(m\sqrt{x^{2}})]\nonumber\\
&  +\frac{mi}{4\pi^{2}\sqrt{-x^{2}}}\theta(-x^{2})K_{1}(m\sqrt{-x^{2}}).
\end{align}
In the neighborhood of the light cone $D^{+}$ has the form%
\begin{equation}
D^{+}(x)\simeq\frac{1}{4\pi}\epsilon(x^{0})\delta(x^{2})+\frac{i}{4\pi
^{2}x^{2}}-\frac{im^{2}}{8\pi^{2}}\ln(\frac{m\sqrt{x^{2}}}{2})-\frac{m^{2}%
}{16\pi}\epsilon(x^{0})\theta(x^{2}).
\end{equation}
To derive the function $F$, it is enough to use only one term%
\begin{equation}
D^{+}(x)\simeq\frac{i}{4\pi^{2}x^{2}},
\end{equation}
with infrared cut-off
\begin{equation}
D^{+}(x)=\frac{i\exp(-\mu x)}{4\pi^{2}x^{2}}.
\end{equation}
Using the formula in the appendix $A$%
\begin{equation}
F=ie^{2}m^{2}\int_{0}^{\infty}\alpha d\alpha D_{+}(x+\alpha r,\mu)-e^{2}%
\int_{0}^{\infty}d\alpha D_{+}(x+\alpha r,\mu)-ie^{2}(d-1)\frac{\partial
}{\partial\mu^{2}}D^{+}(x+\alpha r,\mu),
\end{equation}
we obtain%
\begin{equation}
F=\frac{e^{2}}{4\pi^{2}}+\frac{e^{2}}{16\pi^{2}}[3-d][\ln(\frac{1}{4}\mu
^{2}|x^{2}|)+2C]+\frac{e^{2}}{4\pi^{2}}(\frac{1}{|x|}+\ln(2)),
\end{equation}
where $C$ is a Euler constant.Neglecting regular contribution in the infrared,
we have the spectral function for infinitesimal cut-off $\mu$%
\begin{align}
\rho(p^{2})  &  =\frac{1}{m^{2}}(\frac{m}{\mu})^{\beta}2^{\alpha/\pi}%
e^{\alpha/\pi-C\beta}(s-1+\epsilon)^{\beta-1},\\
\beta &  =\alpha(d-3)/2\pi,
\end{align}
by%
\begin{equation}
\rho(s)=\int_{-\infty}^{\infty}\frac{dt}{2\pi}e^{i(s-1)m^{2}t}e^{F(it)}%
,s=p^{2}/m^{2}%
\end{equation}
with a help of generalized function%
\begin{equation}
e^{i\gamma\pi/2}\frac{1}{2\pi}\int_{-\infty}^{\infty}\frac{dxe^{-ixy}%
}{(x+i\epsilon)^{\gamma}}=\frac{(y+i\epsilon)^{\gamma-1}}{\Gamma(\gamma)}.
\end{equation}
The propagator in momentum space is given%
\begin{align}
G(p)  &  =\frac{Z}{(p^{2}-m^{2})^{1-\beta}},\\
Z  &  =\mu^{-\beta}2^{\alpha/\pi}e^{\alpha/\pi-C\beta}\Gamma(1-\beta),
\end{align}
provided
\begin{equation}
G(p)=\int\frac{ds\rho(s)}{p^{2}-m^{2}s+i\epsilon}.
\end{equation}
Here we return to three dimenisonal case.For three dimensional Minkowski space
,we have
\begin{equation}
F=\frac{-e^{2}}{8\pi m}E_{1}(i\mu t)(itm+1)+\frac{(d+1)e^{2}}{16\pi}%
\frac{e^{-i\mu t}}{\mu}.\nonumber
\end{equation}
Unfortunately we cannot evaluate the spectral function analytically as in four
dimension.Using asymptotic behaviour of the function $E_{1}(z)$ for small$\ z$
and large $z$
\begin{align}
E_{1}(z)  &  =-C-\ln(z),(|z|\ll1),\nonumber\\
E_{1}(z)  &  =\frac{e^{-z}}{z}\{1-\frac{1}{z}+\frac{1\cdot2}{z^{2}}%
-\frac{1\cdot2\cdot3}{z^{3}}+...\}(|arcz|<\frac{3\pi}{2})
\end{align}
we approximate $\exp(F)$ by the rough functional form in the following.In the
gauge $d=-1$ for $\exp(F(t))=e^{C}(\mu it)^{D}$ for $|t|\leq1/\mu$ ,and
$\exp(F)=1$ otherwise for $D=(e^{2}/8\pi m)=1$ and finite $\mu$ case. We
separate the integral in the three regions%
\begin{align}
\rho_{1}(s)  &  =[\int_{-\Lambda}^{-1/\mu}\frac{dt}{2\pi}e^{-i(s-1)m^{2}%
t}dt+\int_{1/\mu}^{\Lambda}\frac{dt}{2\pi}e^{-i(s-1)m^{2}t}dt]+e^{C}%
\int_{-1/\mu}^{1/\mu}\frac{dt}{2\pi}e^{-i(s-1)m^{2}t}i\mu t\\
&  =\frac{\sin((s-1)m^{2}\Lambda)-\sin((s-1)m^{2}/\mu)}{\pi(s-1)m^{2}}%
+e^{C}[\frac{-\cos((s-1)m^{2}/\mu)}{\pi(s-1)m^{2}}+\frac{\mu\sin
((s-1)m^{2}/\mu)}{(s-1)^{2}m^{4}}]\nonumber\\
&  \rightarrow\lbrack\delta((s-1)m^{2})-\frac{\sin((s-1)m^{2}/\mu)}%
{\pi(s-1)m^{2}}]+e^{C}[\frac{-\cos((s-1)m^{2}/\mu)}{\pi(s-1)m^{2}}+\frac
{\sin((s-1)m^{2}/\mu)}{\pi(s-1)^{2}m^{4}/\mu}],\\
\rho_{2}(s)  &  =m\sqrt{s}\rho_{1}(s).
\end{align}
After calculation we have the limit $\Lambda\rightarrow\infty.$From the above
formula we see there is a pole at $s=1$ and that the third and fourth term
cancells at $s=1$.Then these terms\ do not have a singularity at $s=1.$In the
limit $\mu\rightarrow0$,spectral function $\rho$ vanishes.The above result is
one of the main goals in this work.

\subsubsection{perturbative spectral function}

Perturbative $O(e^{2})$ spectral function can be obtained by the usual
definition
\begin{equation}
\rho_{1}(p^{2})=\int d^{3}x\frac{d^{2}r}{(2\pi)^{2}}\frac{d^{2}k}{(2\pi)^{2}%
}\sum_{\lambda,S}T_{1}\overline{T_{1}}(e^{-i(p-k-r)\cdot x}-e^{i(p+k+r)\cdot
x}).
\end{equation}
If we integrate $x$ first,we obtain $(2\pi)^{2}\delta^{(3)}(k+r-p)$ for
energy-momentum conservation.In our case first we integrate $k$.After that we
exponentiate the function $F$ and integrate $r$ in the non perturbative
case.At that stage the results are position dependent.Finaly if we integrate
$x$ we obtain the desired spectral function $\rho_{1}(p^{2})$. Here we
evaluate $F$ from one-photon state marix element $T_{1}$(19) which appeared in
section II A.Thanks to Ward-identity
\begin{align}
\frac{1}{\gamma\cdot(r+k)-m}\gamma\cdot kU(r)  &  =\frac{1}{\gamma
\cdot(r+k)-m}[\gamma\cdot(r+k)-m-(\gamma\cdot r-m)]U(r)\nonumber\\
&  =U(r)
\end{align}
for the $k_{\mu}k_{\nu}$ part of photon polarization sum,we obtain
\begin{equation}
-\sum_{\lambda,S}T_{1}\overline{T_{1}}=e^{2}\frac{m}{E_{r}}\frac{1}{2k_{0}%
}\frac{\gamma\cdot r+m}{2m}[\frac{m^{2}}{(r\cdot k)^{2}}+\frac{1}{r\cdot
k}+\frac{d-1}{k^{2}}].
\end{equation}
Since the propagator is normalized as $(\gamma\cdot r+m)\exp(F)$, we may drop
the $(\gamma\cdot r+m)/2m$ as
\begin{equation}
F=-e^{2}\int\frac{d^{3}ke^{ik\cdot x}}{(2\pi)^{2}}\theta(k^{0})\delta
(k^{2})\frac{m}{E_{r}}[\frac{m^{2}}{(r\cdot k)^{2}}+\frac{1}{(r\cdot k)}%
+\frac{(d-1)}{k^{2}}],
\end{equation}
where we used covariant$\ d$ gauge photon propagator and $\delta(k^{2})$ is
read as the imaginary part of the free photon propagator.%
\begin{equation}
-iD_{0}^{\mu\nu}(k)=\frac{1}{k^{2}+i\epsilon}[g^{\mu\nu}+(d-1)\frac{k^{\mu
}k^{\nu}}{k^{2}}].
\end{equation}
In the appendix $A$ explicit form of the function $F$ is given.In section III
we discuss the position space propagator by using the function $F$. From
(73),(75) in the Feynman gauge we obtain%
\begin{equation}
(2\pi)^{2}\rho^{(2)}(p^{2})\equiv-e^{2}[\frac{4m^{2}}{(p^{2}-m^{2})^{2}}%
+\frac{2}{p^{2}-m^{2}}]\int\frac{d^{2}rd^{2}k}{2E_{r}2k_{0}}\delta
^{(3)}(r+k-p),
\end{equation}
where we omitt the factor $(\gamma\cdot p+m)$.The phase space integral is
evaluated in center-of-mass system,namely $p_{i}=0,p_{0}=p$
\begin{equation}
\int\frac{d^{2}rd^{2}k}{2E_{r}2k_{0}}\delta^{(3)}(r+k-p)=\frac{\pi}{4p_{0}}.
\end{equation}
where%
\begin{align*}
p_{0}  &  =\sqrt{r^{2}+m^{2}}+\sqrt{k^{2}+\mu^{2}},\mathbf{r}+\mathbf{k}=0,\\
\delta(\sqrt{r^{2}+m^{2}}+\sqrt{r^{2}+\mu^{2}}-p_{0})  &  =\frac{1}{f^{\prime
}(r_{0})}\delta(r-r_{0}),\\
\frac{d}{dr}(\sqrt{r^{2}+m^{2}}+r-p_{0})  &  =\frac{r}{\sqrt{r^{2}+m^{2}}%
}+\frac{r}{\sqrt{r^{2}+\mu^{2}}}\\
&  =\frac{r(\sqrt{r^{2}+m^{2}}+\sqrt{r^{2}+\mu^{2}})}{\sqrt{r^{2}+m^{2}}%
\sqrt{r^{2}+\mu^{2}}}=\frac{p_{0}r}{E_{r}E_{k}},\\
r_{0}^{2}  &  =\frac{(p_{0}^{2}-(m+\mu)^{2})(p_{0}^{2}-(m-\mu)^{2})}%
{4p_{0}^{2}}.
\end{align*}
Therefore,we have%
\begin{equation}
\frac{1}{4p_{0}}\int\frac{d\Omega rdr}{r}\delta(r-r_{0})=\frac{\pi}{4p_{0}}.
\end{equation}
where $p^{2}=(r+k)^{2}=s$.Then the spectral function are given
\begin{align}
(2\pi)^{2}\rho_{1}^{(2)}(s)  &  =-e^{2}\frac{\pi}{4\sqrt{s}}[\frac{4m^{2}%
}{(s-m^{2})^{2}}+\frac{2}{s-m^{2}}]\theta(s-m^{2}),\\
\rho_{2}^{(2)}(s)  &  =\sqrt{s}\rho_{1}^{(2)}(s),
\end{align}
where $p^{2}=(r+k)^{2}=s$.Propagator at $O(e^{2})$ is expressed in the Lehmann
representation \qquad%
\begin{equation}
S_{F}^{^{\prime}}(p)=\frac{i(\gamma\cdot p+m)}{p^{2}-m^{2}+i\epsilon}%
-\int_{(m+\mu)^{2}}^{\infty}\frac{i(\gamma\cdot p+\sqrt{s})\rho_{1}%
^{(2)}(s)ds}{p^{2}-s+i\epsilon},
\end{equation}
where $\mu$ is an infrared cut-off. We get%
\begin{align}
\int_{(m+\mu)^{2}}^{\infty}\frac{\gamma\cdot p\rho_{1}^{(2)}(s)ds}%
{p^{2}-s+i\epsilon}  &  =-\frac{e^{2}\gamma\cdot p}{8\pi}[\frac{1}{p^{2}%
-m^{2}}(\frac{1}{\mu}+\frac{1}{2m}-\frac{2}{m}\ln(\frac{2m}{\mu})+\frac
{1}{\sqrt{p^{2}}}\ln(\frac{m-\sqrt{p^{2}}}{m+\sqrt{p^{2}}}))\nonumber\\
&  +\frac{2m}{(m^{2}-p^{2})^{2}}(\ln(\frac{2m}{\mu})+\frac{m}{\sqrt{p^{2}}}%
\ln(\frac{m-\sqrt{p^{2}}}{m+\sqrt{p^{2}}})].
\end{align}%
\begin{align}
\int_{(m+\mu)^{2}}^{\infty}\frac{\rho_{2}^{(2)}(s)ds}{p^{2}-s+i\epsilon}  &
=-\frac{e^{2}}{8\pi}[\frac{1}{p^{2}-m^{2}}(\ln(\frac{m^{2}-p^{2}}{2m\mu
})+\frac{m}{\mu})\nonumber\\
&  +\frac{2m^{2}}{(m^{2}-p^{2})^{2}}\ln(\frac{m^{2}-p^{2}}{2m\mu})].
\end{align}
Next we evaluate the gauge dependent part.
\begin{align}
f(x)  &  \equiv-(d-1)e^{2}\int\frac{d^{3}k\theta(k_{0})\delta(k^{2}-\mu
^{2})e^{ik\cdot x}}{(2\pi)^{2}(k^{2}-\mu^{2})}\nonumber\\
&  =-e^{2}(d-1)\frac{\partial}{\partial\mu^{2}}\frac{e^{-\mu\sqrt{-x^{2}}}%
}{8\pi\sqrt{-x^{2}}}=\frac{e^{2}(d-1)e^{-\mu\sqrt{-x^{2}}}}{16\pi\mu}.
\end{align}
First we consider the $d=-1$ gauge case.Following (73) spectral function is
directly given as%
\begin{align}
&  \int d^{3}x\int\frac{d^{3}r}{(2\pi)^{3}}\epsilon(r_{0})\delta(r^{2}%
-m^{2})e^{-i(p-r)\cdot x}\frac{-e^{2}e^{-\mu|x|}}{8\pi\mu}\nonumber\\
&  =\int d^{3}xe^{-ip\cdot x}\frac{e^{2}e^{-(m+\mu)|x|}}{4\pi|x|8\pi\mu}%
=\frac{e^{2}}{8\pi\mu}\frac{1}{p^{2}+(m+\mu)^{2}}.
\end{align}
In Minkowski space we have
\begin{equation}
\rightarrow\frac{e^{2}}{8\pi\mu}\frac{-1}{p^{2}-m^{2}+i\epsilon}=\frac{e^{2}%
}{8\pi\mu}(-P(\frac{1}{p^{2}-m^{2}})+i\pi\delta(p^{2}-m^{2})).
\end{equation}
Linear infrared divergence of wave function correction in (84),(85) cancells
with this term.Therefore in $d=-1$ gauge linear divergence is absent.Adding
this term to spectral function we may write%
\begin{equation}
(2\pi)^{2}\rho^{(2)}(s)=-e^{2}\frac{\pi}{4\sqrt{s}}[\frac{4m^{2}}%
{(s-m^{2})^{2}}+\frac{2}{s-m^{2}}]\theta(s-m^{2})+\frac{(2\pi)^{2}(d-1)e^{2}%
}{16\pi\mu}\delta(s-m^{2}).
\end{equation}
In $d=-1$ gauge logarithmic infrared divergence $\ln(2m/\mu)$ can be absorved
to the wave function renormalization constant $Z_{2}.$We have $Z_{2}%
=1-e^{2}/16\pi m.$

\subsection{\bigskip Photon}

For unquenched case we use the dressed photon with fermion loop with $N$
flavours.Spectral functions for dressed photon are given by vacuum
polarization [6]%
\begin{align}
\Pi_{\mu\nu}(k)  &  \equiv ie^{2}N\int\overline{d^{3}}pTr(\gamma_{\mu}\frac
{1}{\gamma\cdot p-m}\gamma_{\nu}\frac{1}{\gamma\cdot(p-k)-m})\nonumber\\
&  =-e^{2}N\frac{T_{\mu\nu}}{8\pi}[(\sqrt{k^{2}}+\frac{4m^{2}}{\sqrt{k^{2}}%
})\ln(\frac{2m+\sqrt{k^{2}}}{2m-\sqrt{k^{2}}})-4m],\\
T_{\mu\nu}  &  =(g_{\mu\nu}-\frac{k_{\mu}k_{\nu}}{k^{2}}),\overline{d^{3}%
}p=\frac{d^{3}p}{(2\pi)^{3}},\nonumber
\end{align}%
\begin{equation}
D_{\mu\nu}^{-1}(k)\equiv T_{\mu\nu}(k^{2}+dk_{\mu}k_{\nu})-\Pi_{\mu\nu}(k).
\end{equation}
Polarization function $\Pi(k)$ is
\begin{align}
\Pi(k)  &  =\frac{-e^{2}}{8\pi}N[(\sqrt{k^{2}}+\frac{4m^{2}}{\sqrt{k^{2}}}%
)\ln(\frac{2m+\sqrt{k^{2}}}{2m-\sqrt{k^{2}}})-4m],\nonumber\\
&  =\frac{-e^{2}}{8}Ni\sqrt{k^{2}}(k^{2}>0,m=0),\nonumber\\
&  =\frac{-e^{2}N}{6\pi m}k^{2}+O(k^{4})(k^{2}/m\ll1).
\end{align}
Fermion mass is assumed to be generated dynamically.In quenched case it is
shown that $m$ is proportional to $e^{2}$[4].For massless case we have for
number of $N$ fermion flavour
\begin{equation}
\rho_{\gamma}^{D}(k)=\frac{1}{\pi}\operatorname{Im}D_{F}(k)=\frac{c\sqrt
{k^{2}}}{k^{2}(k^{2}+c^{2})},c=\frac{e^{2}N}{8},
\end{equation}%
\begin{equation}
D_{\mu\nu}(k)=-\int_{0}^{\infty}\frac{(g_{\mu\nu}-k_{\mu}k_{\nu}/k^{2}%
)\rho_{\gamma}(s)ds}{k^{2}-s+i\epsilon}-d_{R}\frac{k_{\mu}k_{\nu}}%
{(k^{2}+i\epsilon)^{2}},
\end{equation}
where $d_{R}=Z_{3}^{-1}d.$If we include massive fermion loop to photon
spectral function $\rho_{\gamma}$ we have
\begin{equation}
\rho_{\gamma}(s)=\frac{1}{\pi}\operatorname{Im}D_{F}(s)=Z_{3}\delta
(s)+\sigma(s)\theta(s-4m^{2}),
\end{equation}
where $\sigma$ is a imaginary part of the vacuum polarization function
$\Pi(k^{2})$ for $k^{2}\geq4m^{2}.$However real part of vacuum
polarization\ $\Pi(k^{2})$ affects the residue of the massless pole.%
\begin{equation}
D_{\mu\nu}^{F}(k)^{\prime}=\frac{-(g_{\mu\nu}-k_{\mu}k_{\nu}/(k^{2}%
+i\epsilon))}{k^{2}-\Pi(k)+i\epsilon}-d_{R}\frac{k_{\mu}k_{\nu}}%
{(k^{2}+i\epsilon)^{2}}%
\end{equation}
So that the renormalization constant $Z_{3\text{ \ }}$is defined by expanding
$\Pi(k^{2})$ at $k^{2}=0$
\begin{align}
\lim_{k^{2}\rightarrow0}D_{F}(k)  &  =\lim_{k^{2}\rightarrow0}\frac{-1}%
{k^{2}-\Pi(k^{2})+i\epsilon}=\frac{-Z_{3}}{k^{2}+i\epsilon},\\
Z_{3}^{-1}  &  =1-\frac{\Pi(k^{2})}{k^{2}}|_{k^{2}=0}.
\end{align}%
\begin{align}
1  &  =Z_{3}+\int_{4m^{2}}^{\infty}\sigma(s)ds,\\
\sigma(s)  &  =\frac{\operatorname{Im}\Pi(s)}{\pi(-s+\operatorname{Re}%
\Pi(s))^{2}+(\operatorname{Im}\Pi(s))^{2}}\theta(s-4m^{2}).
\end{align}

\section{Analysis in position space}

\subsection{Quenched case}

To evaluate the function $F$ it is helpful to use the exponential
cut-off(infrared cut-off)[3,4].In the appendices the way to evaluate spectral
function $F$ of one-photon state\ in the covariant $d$ gauge is given.%
\begin{equation}
F=\frac{-e^{2}}{8\pi m}(m|x|+1)E_{1}(\mu|x|)+\frac{(d+1)e^{2}e^{-\mu|x|}%
}{16\pi\mu},
\end{equation}
where\ $|x|=\sqrt{-x^{2}}$ and $\mu$ is a bare photon mass and
\begin{equation}
E_{1}(x)=\int_{x}^{\infty}\frac{e^{-t}}{t}dt.
\end{equation}
Short distance behaiviour of $F$ has the following form
\begin{equation}
E_{1}(\mu|x|)\sim-\gamma-\ln(\mu|x|)+\mu|x|,
\end{equation}%
\begin{equation}
F\sim\frac{(d+1)e^{2}}{16\pi\mu}+\frac{e^{2}}{8\pi m}[-\mu|x|+(1+m|x|)(\gamma
+\ln(\mu\left\vert x\right\vert ))]-\frac{(d+1)e^{2}}{16\pi}|x|,(\mu|x|\ll1).
\end{equation}
Long distance behaviour is given by the asymptotic expansion of $E_{1}%
(\mu|x|)$%
\begin{equation}
E_{1}(z)\sim\frac{e^{-z}}{z}\{1-\frac{1}{z}+\frac{1\cdot2}{z^{2}}-\frac
{1\cdot2\cdot3}{z^{3}}+...\},(|\arg z|<\frac{3}{2}\pi),
\end{equation}
therefore we have%
\begin{align}
F  &  \sim\frac{-e^{2}}{8\pi m}(m|x|+1)\frac{e^{-\mu|x|}}{\mu|x|}(1-\frac
{1}{\mu|x|})+\frac{(d+1)e^{2}e^{-\mu|x|}}{16\pi\mu},\\
&  =-\frac{e^{2}}{8\pi\mu}e^{-\mu|x|}+\frac{(d+1)e^{2}e^{-\mu|x|}}{16\pi\mu
},(1\ll\mu|x|),
\end{align}
where $\gamma$ is an Euler constant.In (104) linear term in $\left\vert
x\right\vert $ is understood as the finite mass shift from the form of the
propagator in position space, and $\left\vert x\right\vert \ln(\mu\left\vert
x\right\vert )$ term is a position dependent mass
\begin{align}
m_{S}  &  =m+\frac{(d+1)e^{2}}{16\pi},\\
m(x)  &  =m-\frac{e^{2}}{8\pi}\ln(\mu\left\vert x\right\vert ),
\end{align}
which has mass changing effects at short distance.We have seen the drastic
change of the function $F(\mu,x)$ from short distance to long distance.For
long distance $F$ vanishes and it may be a free particle for fixed $\mu$.The
electron propagator in position space can be written approximately for
arbitrary $D$%
\begin{align}
S_{F}^{^{\prime}}(x)  &  =-(i\gamma\cdot\partial+m)\frac{1}{4\pi\sqrt{x^{2}}%
}\left(
\begin{array}
[c]{c}%
e^{-m|x|}A(\mu|x|)^{D+C|x|}(\mu|x|\leqslant1)\\
e^{-m|x|}(1\eqslantless\mu|x|)
\end{array}
\right)  ,\nonumber\\
&  =(S_{V}(x)+S_{S}(x)),
\end{align}
where%
\begin{equation}
A=\exp(\frac{(d+1)e^{2}}{16\pi\mu}),C=\frac{e^{2}}{8\pi},D=\frac{e^{2}}{8\pi
m}.
\end{equation}
Since $\lim_{x\rightarrow0}|x|^{|x|}=1,$we see that $C|x|$ term dose not
affect the short distance behavior.$A$ is a violent linear infrared divergent
factor.Hereafter we choose $d=-1$ gauge to avoid linear infrared divergence.In
this gauge, for $D>1,$in the limit $\mu\rightarrow0$ for finite $|x|$ scalar
part of the propagator $S_{F}^{\prime}(x)$ vanishes as
\begin{align}
\frac{me^{-m|x|}}{4\pi\sqrt{x^{2}}}(\mu|x|)^{D+C|x|}  &  \rightarrow
0(\mu|x|\ll1),\nonumber\\
\frac{me^{-m|x|}}{4\pi\sqrt{x^{2}}}\exp(-\frac{e^{2}}{8\pi\mu}e^{-\mu|x|})  &
\rightarrow0(1\ll\mu|x|).
\end{align}
Therefore fermion is confined in quenched case for $D>1$.There is no way to
determine the value $D$ in principle.So that we need extra physical condition
such as the finiteness of the vacuum expectation value of chiral order
parameter which is proportional to $-itr(S_{F}^{^{\prime}}(x))$ in the case of
dynamical chiral symmetry breaking.In this sense we see that $S_{F}^{\prime
}(0)$ is finite and the physical mass equals to $m=e^{2}/8\pi$ for $D=1$ with
finite value of $\mu$.In the next section we consider the way to avoid the
vanishment of the propagator by including vacuum ploarization.We will discuss
unquenched case with finite $\mu$ in numerical analysis in Minkowski space in
the next section.

\subsection{ Unquenched case}

In our approximation quenched propagator vanishes in the limit of zero bare
photon mass $\mu.$So that we may apply the spectral function of photon
$\rho_{\gamma}(\mu)$ to evaluate the unquenched fermion proagator.In that case
we simply integrate the function $e^{F(x,\mu)}\rho_{\gamma}(\mu)$ for $\mu
$,where $\mu$ is a invariant mass of two fermion-antifermion intermediate
state.Spectral function of photon with massless fermion loop is derived from
(92) and we have
\begin{equation}
\rho_{\gamma}^{0}(\mu)=\frac{2c}{\pi(\mu^{2}+c^{2})},c=\frac{e^{2}}{8}%
,Z_{3}=\int_{0}^{\infty}d\mu\rho_{\gamma}^{0}(\mu)=1,
\end{equation}
where photon has not a simple pole at $k^{2}=0.$In this case the spectral
function of the fermion in position space is given
\begin{equation}
\widetilde{\rho}(x)=\int_{0}^{\infty}d\mu\rho_{\gamma}^{0}(\mu)e^{F(x,\mu)}.
\end{equation}
However massless loop are not able to suppress large $\mu$ region as
$1/\mu^{3}$.In fact integral which appears at short distance part of
$\exp(F(x,\mu)$%
\begin{equation}
\int_{0}^{\infty}\rho_{\gamma}^{0}(\mu)(\mu|x|)^{D}d\mu
\end{equation}
is logarithmically divergent for $D=1$.So that we may consider spectral
function with the massive fermion loop.

In terms of real and imagainary part of polarization function $\Pi(k)$%
\begin{align}
\Im\Pi(k)  &  =-\frac{e^{2}}{8}[\sqrt{k^{2}}+\frac{4m^{2}}{\sqrt{k^{2}}%
}]\theta(k^{2}-4m^{2}),\\
\Re\Pi(k)  &  =-\frac{e^{2}}{8\pi}[(\sqrt{k^{2}}+\frac{4m^{2}}{\sqrt{k^{2}}%
})\ln|\frac{2m+\sqrt{k^{2}}}{2m-\sqrt{k^{2}}}|-4m].
\end{align}
spectral function of photon is given.%
\begin{equation}
\rho_{\gamma}(\mu^{2})=\frac{-\Im\Pi(\mu)}{\pi\lbrack(\mu^{2}-\Re\Pi(\mu
))^{2}+(\Im\Pi(\mu)^{2})}.
\end{equation}
This spectral function damps as $1/\mu^{3}$ at large $\mu$ by (100) in section
II.C.Small $|x|$ region of the fermion propagator is modified with $N_{F}$
massive fermion loop for photon spectral function
\begin{align}
S_{F}^{^{\prime}}(x)  &  =-(i\gamma\cdot\partial+m)\frac{e^{-m\sqrt{-x^{2}}}%
}{4\pi\sqrt{x^{2}}}\widetilde{\rho}(x),\\
\widetilde{\rho}(x)  &  =\int_{4m^{2}}^{\infty}d\mu^{2}\sigma(\mu
^{2})e^{F(x,\mu)},
\end{align}
where all bare photon propagators in $\exp(F)$ are dressed by $\rho_{\gamma}%
.$The above result is the second goal in our work.In the massive fermion case
renormalization constant $Z_{3}$ is given by equations (92) and (98) in
section II.C
\begin{equation}
Z_{3}^{-1}=(1-\frac{\Pi(k)}{k^{2}}|_{k^{2}=0})=(1+\frac{Ne^{2}}{6\pi m}%
)=\frac{7}{3}(N=1)
\end{equation}
with our choice of physical mass $m=e^{2}/8\pi.$In Fig.1 we see fermion loop
effects lead infrared finite spectral function $\widetilde{\rho}(x)$ with
dynamical fermion mass $m=\alpha/N$ $,e^{2}=\alpha/N$ from $N=1$(lower) to
$3$(upper) with $d=-1,\alpha=e^{2}N/8\pi$ is fixed to unity$.$%
\begin{figure}
[ptb]
\begin{center}
\includegraphics[
height=2.5097in,
width=2.5097in
]%
{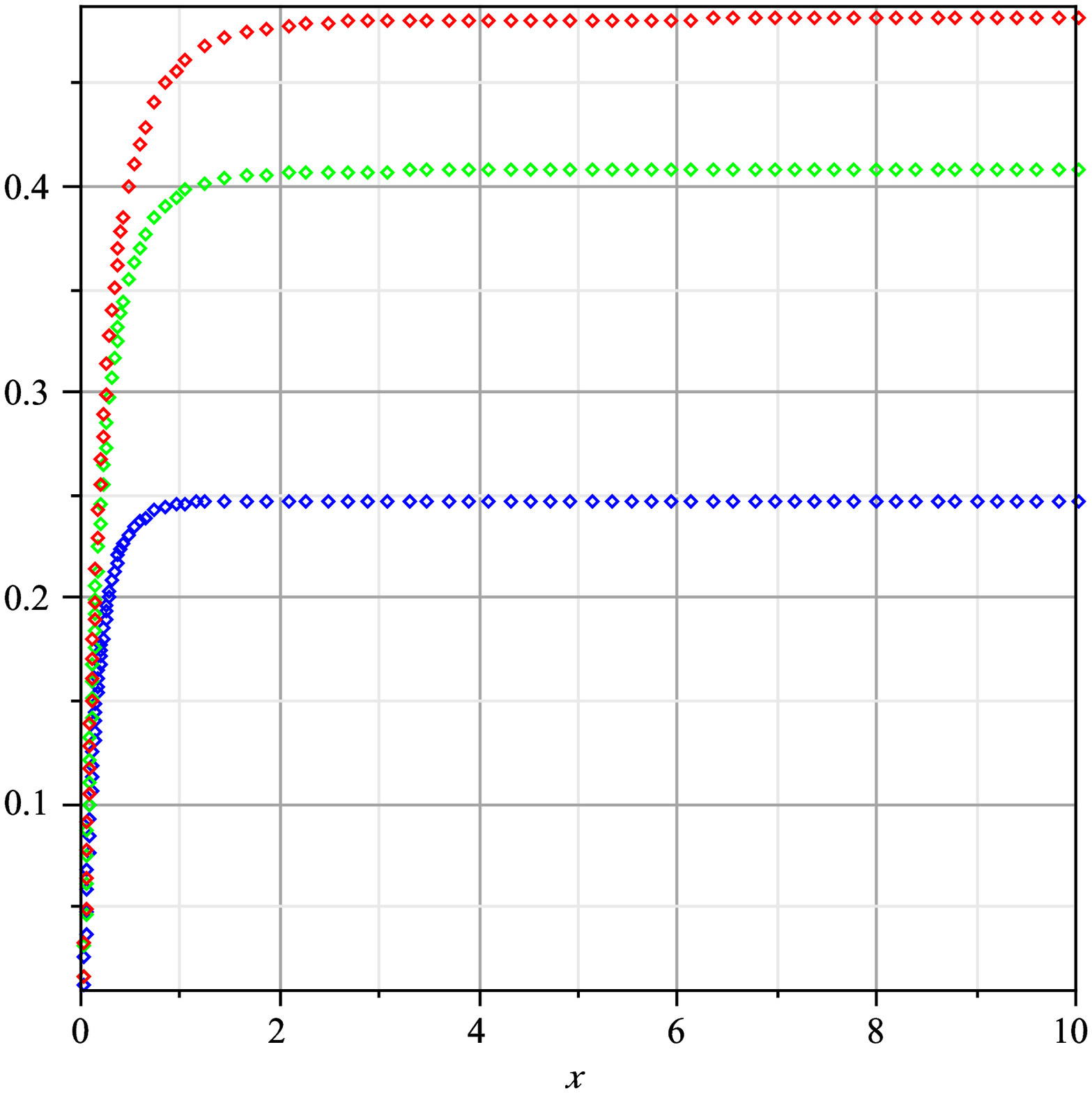}%
\caption{$\widetilde{\rho}(x)$ for $\alpha=1,N=1(blue)..3(red).$}%
\label{1}%
\end{center}
\end{figure}
Next we evaluate the spectral function in $d=-1$ gauge.%
\begin{equation}
F=\frac{-e^{2}}{8\pi m}E_{1}(i\mu t)(itm+1).
\end{equation}%
\begin{equation}
\rho_{1}(s)=\int_{-\infty}^{\infty}\frac{dt}{2\pi}e^{-i(s-1)m^{2}%
t-\epsilon|t|}\int_{4m^{2}}^{\infty}d\mu^{2}\sigma(\mu^{2})e^{F(i\mu
t)},s=p^{2}/m^{2}.
\end{equation}
In Fig.2 we show $\rho_{1}(s)$.At $s=1$ there seems to be a pole like
singularity which is expected by long distance behavior of the function
$e^{F}\simeq1$.The hight of the peak at $s=1$ is consistent with $\Lambda/\pi$
from the approximate formula $\ \delta(s-1)=\sin(\Lambda(s-1))/\pi(s-1)$,
$\Lambda=6.17\times10^{5}.$
\begin{figure}
[ptb]
\begin{center}
\fbox{\includegraphics[
trim=0.000000in 0.000000in 0.414602in -0.648255in,
height=2.5097in,
width=2.5097in
]%
{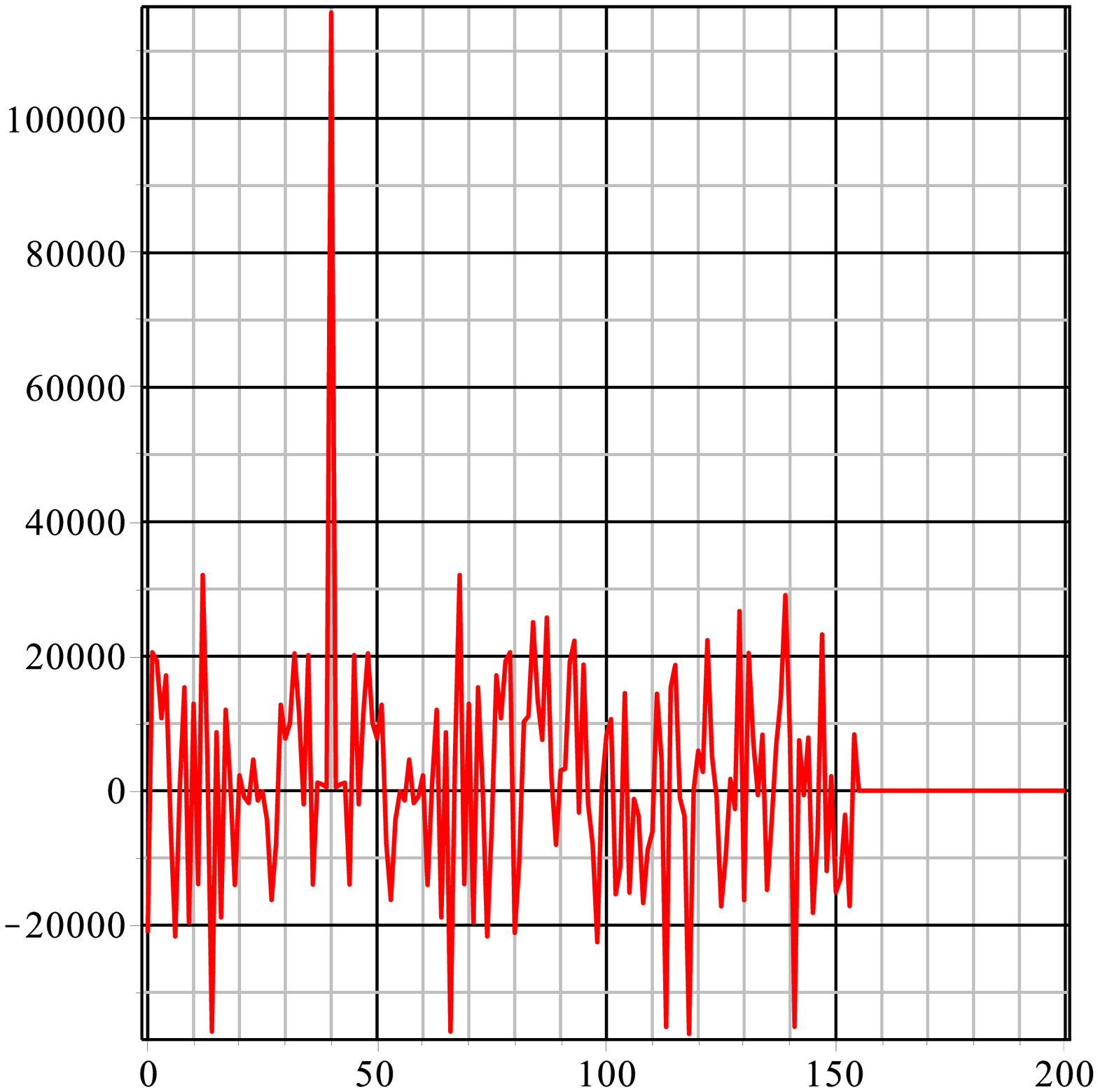}%
}\caption{$\rho_{1}(s)$ for $e^{2}=1,N=1,s=.025l.$}%
\label{2}%
\end{center}
\end{figure}
$\ \ $

\section{Renormalization constant and order parameter}

\subsection{\bigskip by Spectral function}

In this section we consider the renormalization constant and chiral order
parameter in our model.It is easy to evaluate the renormalization constant by
the equation
\begin{equation}
\lim_{\gamma\cdot p\rightarrow m}-iS_{F}^{^{\prime}}(p)=\frac{Z_{2}}%
{\gamma\cdot p-m}.
\end{equation}
where $Z_{2\text{ }}$is defined for one particle state in the theory.In
qenched approximation with finite cut-off we have a $\delta(s-1)$ term in
$\rho_{1}$ which implies
\begin{equation}
Z_{2}\neq0.
\end{equation}
Formally renormalization constant $Z_{2}$ is defined in our approximation%
\begin{align}
Z_{2}  &  =\int_{1}^{\infty}m^{2}\rho_{1}(s)ds,\\
\int_{1}^{\infty}m^{2}\rho_{1}(s)ds  &  =\int_{-\infty}^{\infty}\frac{m^{2}%
dt}{2\pi}\int_{1}^{\infty}dse^{-i(s-1)m^{2}t-\epsilon s}\int_{0}^{\infty}%
\rho_{\gamma}(\mu^{2})d\mu^{2}e^{F(i\mu t)}\nonumber\\
&  =\int_{-\infty}^{\infty}\frac{dt}{2\pi}\frac{-e^{-i(s-1)m^{2}t-\epsilon s}%
}{it}|_{s=1}^{s=\infty}\int_{0}^{\infty}\rho_{\gamma}(\mu^{2})d\mu
^{2}e^{F(i\mu t)}\nonumber\\
&  =\int_{4m^{2}}^{\infty}d\mu^{2}\sigma(\mu^{2})\int_{-\infty}^{\infty}%
\frac{dt}{2\pi}\frac{e^{F(i\mu t)}}{it},
\end{align}
where we add the convergence factor $\exp(-\epsilon s)$ to define the definite
integral with infinitesimal $\epsilon$.Following the above formula we
evaluated $Z_{2}$ numerically for $d=-1$ gauge with $e^{2}=1.$We have
$Z_{2}=0.283,0.400$ for $N=1$ and $3$ respectively.If we \ consider $N_{F}$
flavor of massless four component fermion,there is a chiral like symmetry
$U(2N_{F})$ which breaks down to $SU(N_{F})\otimes SU(N_{F})\otimes
U(1)\otimes U(1)$ by fermion mass generation[9,10].Chiral like symmery is
realized by $\gamma_{3},\gamma_{5}$ transformation in place of $\gamma_{5}$ in
four dimension.There appears a pair of Goldstone boson like Pion in QCD.Chiral
order parameter for each flavour $\left\langle \overline{\psi}\psi
\right\rangle $ is given
\begin{equation}
\left\langle \overline{\psi}\psi\right\rangle =-itrS_{F}^{^{\prime}}(x).
\end{equation}
In position space we evaluate directly from (119),(120)%
\begin{equation}
\left\langle \overline{\psi}\psi\right\rangle =-4\lim_{|x|\rightarrow0}%
\int_{2m}^{\infty}d\mu\frac{\alpha e^{-m|x|}e^{\widetilde{F}(x,\mu)}}%
{N\cdot8\pi|x|},
\end{equation}
where $m=\alpha/N$.Since $S_{F}^{\prime}(x-x^{\prime})$ may be expressed as
the sum of even and odd function in $(x-x^{\prime})$[7],odd part vanishes at
$x=x^{\prime}.$Then we have a half value of the propagator at the origin as
the vacuum expectation value. From the above equation we see that
$\left\langle \overline{\psi}\psi\right\rangle $ $\neq0$ and is finite only if
$D=\alpha/Nm=1$.At least for weak coupling we get the value of order parameter
$\left\langle \overline{\psi}\psi\right\rangle $ as $-3.26\times10^{-3}e^{4}$
for $N=1,e^{2}=1$ case.We used infrared cut-off for quenched part $\mu=.05$
which is smaller than $2m_{f}=1/4\pi=.079$.Contribution of quenched part to
the condensation is $\left\langle \overline{\psi}\psi\right\rangle
_{Q}=-.56\times10^{-3}e^{4}.$This is the third goal in our work.For large
$N,m$ is $O(\alpha/N)$ and mass generation is suppressed which are shown in
Fig.3.In Fig.4 coupling dependence of order parameter at fixed $N$ is
shown.Since ferimon mass is proportional to $\alpha$,fermion loop effect is
suppresed at large $\alpha$.There may be a critical value of $\alpha.$%

\begin{figure}
[ptb]
\begin{center}
\includegraphics[
height=2.5097in,
width=2.5097in
]%
{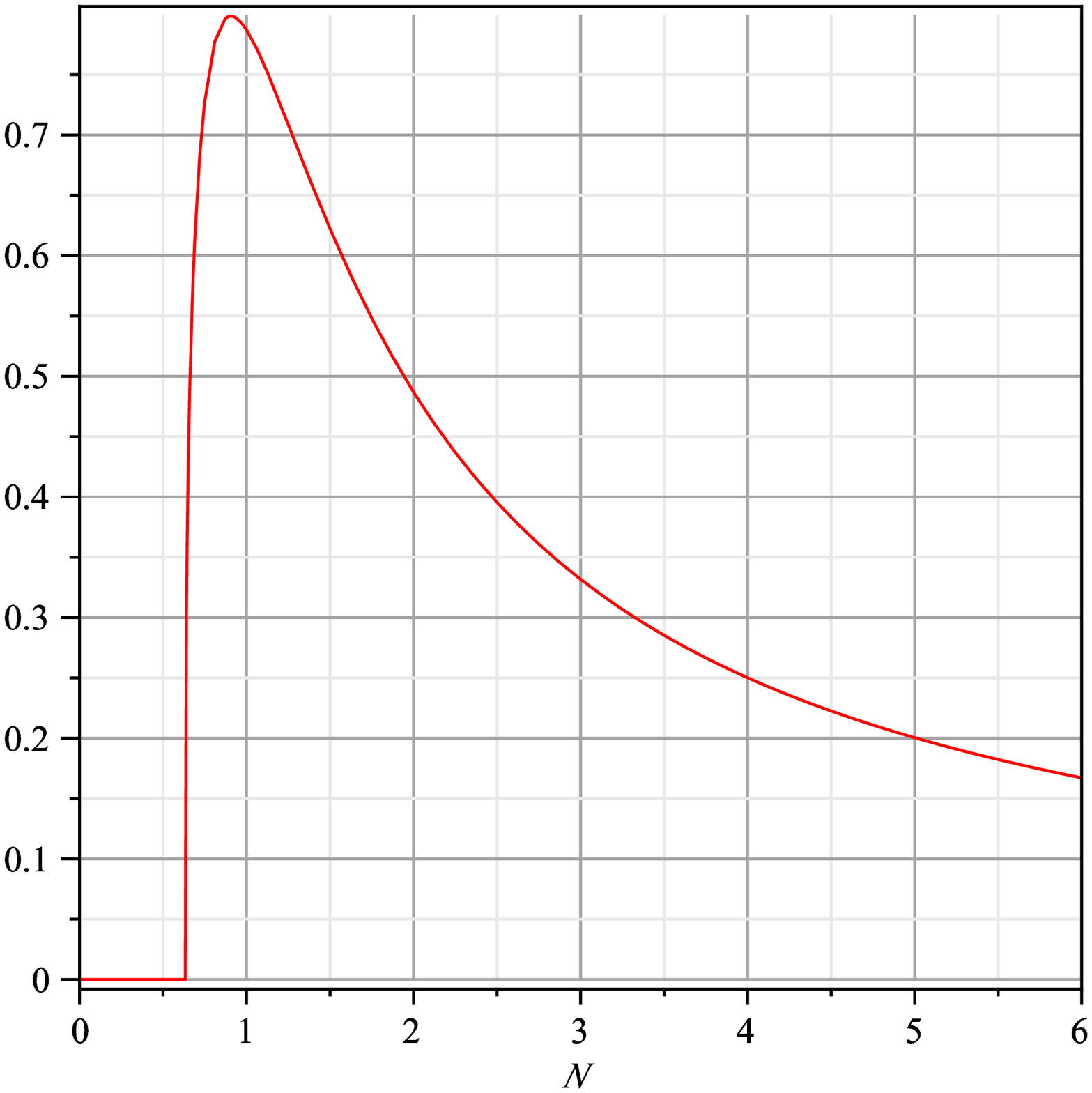}%
\caption{$-\left\langle \overline{\psi}\psi\right\rangle $ for $N,\alpha=1.$}%
\end{center}
\end{figure}
%

\begin{figure}
[ptb]
\begin{center}
\includegraphics[
height=2.5097in,
width=2.5097in
]%
{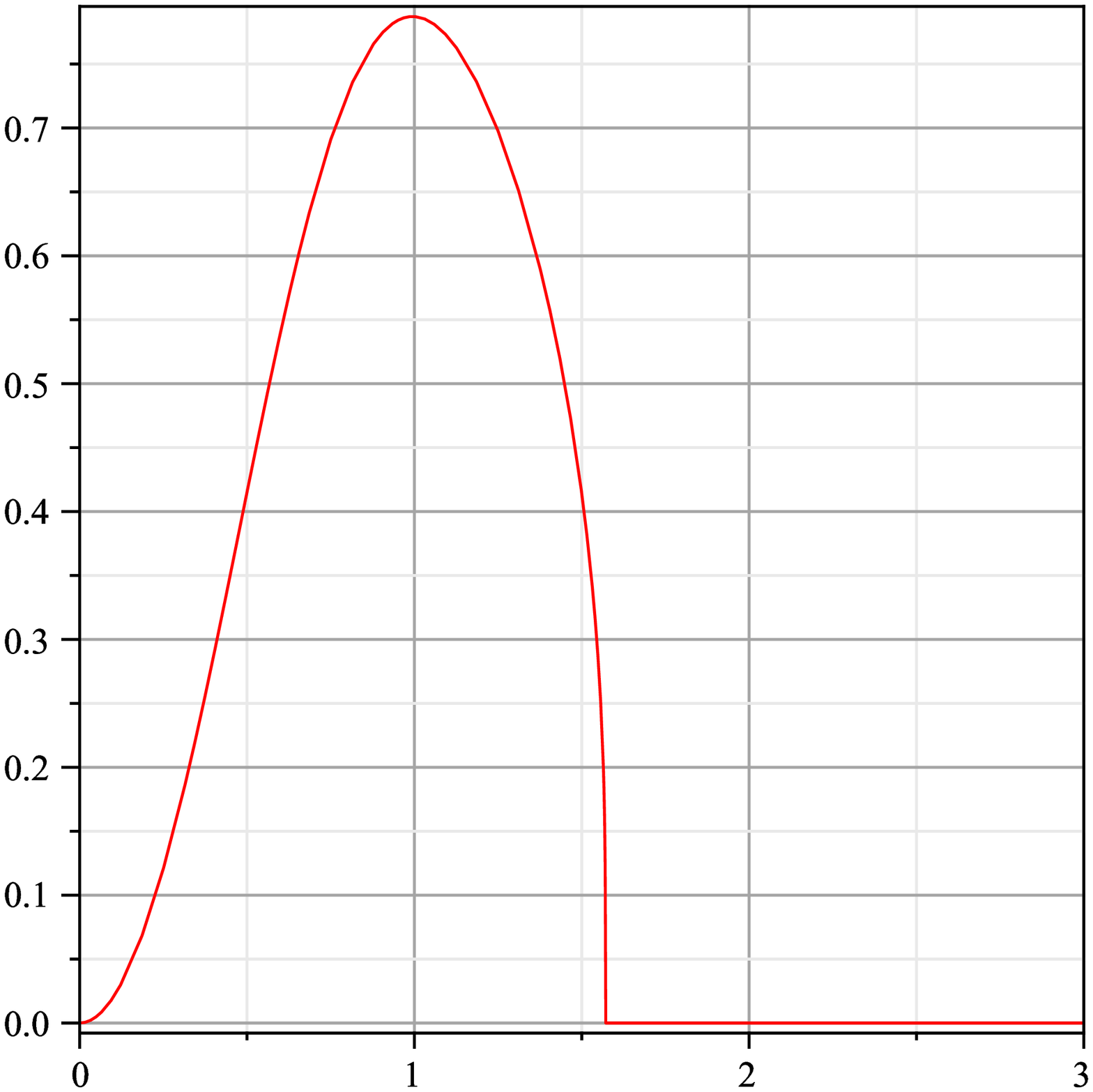}%
\caption{$-\left\langle \overline{\psi}\psi\right\rangle $ for $\alpha,N=1.$}%
\end{center}
\end{figure}

\subsection{by Dyson-Schwinger equation}

We have a similar solution of the propagator at short distance which is known
by the analysis of Dyson-Schwinger equation[10].The value of order parameter
$\left\langle \overline{\psi}\psi\right\rangle $ as $-3.26\times10^{-3}e^{4}%
$,$N=1$,$d=-1$ case for weak coupling may be compared with $-3.21\sim
-3.5\times10^{-3}e^{4}$ for quenched case with the gauge parameter range
$d=0\sim1$ and vertex correction of Ball-Chiu(BC) or Curtis-Penningtong(CP)
ansatz for the transverse vertex.The BC vertex ansatz which satisfies
Ward-Takahashi-identity%
\begin{equation}
(p-q)_{\mu}\Gamma_{\mu}(p,q)=S_{F}^{-1}(q)-S_{F}^{-1}(p)
\end{equation}
is known as
\begin{align}
\Gamma_{\mu}^{BC}(p,q)  &  =\frac{A(p^{2})+A(q^{2})}{2}\gamma_{\mu}%
+\frac{\Delta A}{2}(p+q)\cdot\gamma(p+q)_{\mu}-\Delta B(p+q)_{\mu},\\
\Delta A  &  =\frac{A(p^{2})-A(q^{2})}{p^{2}-q^{2}},\Delta B=\frac
{B(p^{2})-B(q^{2})}{p^{2}-q^{2}}.
\end{align}
The Dyson-Schwinger equation for the fermion propagator with BC vertex ansatz
is written in the Landau gauge%

\begin{equation}
A(p)=1-\frac{e^{2}}{4\pi^{2}p^{2}}\int_{0}^{\infty}\frac{dqq^{2}}%
{A(q)^{2}q^{2}+B(q)^{2}}[\Delta A(p,q)A(q)+\Delta B(p,q)B(q)]I_{1}(p,q),
\end{equation}%
\begin{equation}
B(p)=\frac{e^{2}}{4\pi^{2}}\int_{0}^{\infty}\frac{dqq^{2}}{A(q)^{2}%
q^{2}+B(q)^{2}}[(A(p)+A(q))B(q)I_{0}(p,q)+(\Delta A(p,q)B(q)-\Delta
B(p,q)A(q))I_{1}(p,q)],
\end{equation}
where
\begin{align}
I_{0}(p,q)  &  =-\frac{1}{2pq}\ln(\frac{(p-q)^{2}}{(p+q)^{2}}),\\
I_{1}(p,q)  &  =-\frac{p^{2}+q^{2}}{2pq}\ln(\frac{(p-q)^{2}}{(p+q)^{2}}-2.
\end{align}
Here we notice that if we drop the $\Delta B$ term in the BC vertex,value of
order parameter is equal to the case of bare vertex in the Landau
gauge$(2.1\times10^{-3}e^{4})$ and we have $A(p)\simeq1$ except for the low
momentum region.In the chiral symmetric case with $B(p)=0$,thanks to
Ward-Takahashi identity we have $A(p)=1.$With BC vertex the value of
$\left\langle \overline{\psi}\psi\right\rangle $ is larger by $(1.6-1.8)$
times than that for bare vertex in the Dyson-Schwinger equation.

\section{Summary}

We evaluated the fermion propagator in three dimensional QED with dressed
photon by the dispersion method and Dyson-Schwinger equation with vertex
correction.In the $O(e^{2})$ fermion position space spectral function $F$ we
obtain finite mass shift,wave function renormalization and position dependent
mass.However there remains infrared divergences such as linear and logarithmic
ones which were regularized by bare photon mass $\mu.$In our approximation the
linear divergence is absent in covariant $d=-1$ gauge.After exponentiation of
$F$ we have position space spectral function proportional to $(\mu\sqrt
{-x^{2}})^{\beta}$ for short distance and $\exp(-e^{2}/(8\pi\mu)\exp(-\mu
\sqrt{-x^{2}})$ ) for long distance$,\beta=e^{2}/8\pi m.$These facts imply
fermion to be free at low energy.This feature is not given in the
Dyson-Schwinger equation.In the limit $\mu\rightarrow0,$the quenched spectral
function vanishes for both short and long distance.If we include massive
fermion loop to vacuum polarization for photon,imaginary part of the photon
propagator gives us a modified photon spectral function $\rho(\mu^{2})$ in
place of quenched $\delta(\mu^{2})$ function and we may avoid the vanishment
of the propagator in unquenched case.Our analysis is an extension of
dispersion like method applied in the determination of one-particle
singularity in scalar QED[3].In Minkowski space the spectral function of
fermion is evaluated numerically and $\rho_{1}(s)$ shows the simple pole
structure near $p^{2}=m^{2}$.This fact is consistent with a finite
renormalization constant $Z_{2}$ that was obtained in the previous section.In
our approximation only short distance behavior is modified in position
space.So that electron is free at low energy which is suitable for the
application of the model in the condensed matter physics in (2+1) dimension.It
has been shown that there is no infrared divergences in non-covariant
gauges[3].In (2+1)-dimensional case our approximation leads results similar to
non-covariant gauge case in the infrared,where there may be no unphysical
degrees of freedom for photon polarization.As far as we know only Gauge
Technique allows us to have a confining solution with infrared anomalous
dimension provided by massless loop[6].Generally speaking propagator or
self-energy obtained by Dyson-Schwinger is gauge dependent especially in the
infrared.So that we have not a definite conclusion about confinement or
existence of asymptotic field.On the other hand our approximation with
soft-photon exponentiation is gauge invariant in the infrared,it strongly
suggests the existence of asymptotic fields in\ (2+1)-dimension.For the case
of chiral symmetry breaking,if we set the anomalous dimension to be unity at
short distance we obtain finite vacuum expectation value $\left\langle
\overline{\psi}\psi\right\rangle .$This value agrees quite well with that
provided by quenched Dyson-Schwinger equation.In our approximation vacuum
expectation value and physical mass are not so sensitive to flavor number $N$
which may be linear in $1/N$ in comparison with Dyson-Schwinger analysis with
massless fermion loop.In the strong coupling with fixed $N$ order parameter
vanishes by suppression of heavy fermion loop,where fermion mass is
proportional to $e^{2}$.If we neglect fermion mass our results may reproduce
the results in an earlier work of Templeton,where fermion mass is assumed to
be neglected at high temperature[2].

\section{Acknowledgement}

The author would like to thank to Prof.Robert Delbourgo at University of
Tasmania for his introduction of Gauge Technique.He also thanks to Prof.Roman
Jackiw at MIT to recommend to apply his dispersion like method to our problem.

\section{References\newline}

\noindent\lbrack1]R.Jackiw,S.Templeton,Phys.Rev.D.\textbf{23}%
(1981)2291.\noindent{}

\noindent\nolinebreak{}[2]Stephen.Templeton,Phys.Rev.D.24(1981)3134.\newline%
[3]R.Jackiw,L.Soloviev,Phys.Rev.\textbf{137}%
.3(1968)1485;S.Weinberg,Phys.Rev.\textbf{140}.2B(1965)516.\newline[4]Yuichi
Hoshino,\textbf{JHEP0409}:048,2004.\newline[5]K.Nishijima,\textbf{Fields and
Particles},W.A.BENJAMIN,INC(1969).\newline%
[6]A.B.Waites,R.Delbourgo,Int.J.Mod.Phys.\textbf{A7}(1992)6857.\newline%
[7]James D. Bjorken and Sidney D.Drell,\textbf{Relativistic Quantum
Fields},McGraw-Hill Book Company.\newline[8]N.N.BOGOLIUBOV and
D.V.SHIRKOV,\textbf{INTRODUCTION TO THE THEORY OF QUANTIZED} \textbf{FIELDS}

,WILEY-INTERSCIENCE.\newline[9]C.J.Burden,Nuclear Physics \textbf{B}
387(1992)419-446.\newline%
[10]C.S.Fischer,R.Alkofer,T.Dahm,P.Maris,Phys.Rev.\textbf{D70}%
,073007(2004):[arXiv:hep-th/0407014].\newline

\section{Appendices}

\subsection{ Evaluation of specral function of one-photon state}

In this section we evaluate the $O(e^{2})$ spectral function $F$%
\begin{equation}
F=-e^{2}\int\frac{d^{3}k}{(2\pi)^{2}}e^{ik\cdot x}\theta(k^{0})\delta
(k^{2})[\frac{m^{2}}{(r\cdot k)^{2}}+\frac{1}{(r\cdot k)}+\frac{(d-1)}{k^{2}%
}].
\end{equation}
The gauge dependent term is\ an off-shell quantity in general.Here we
introduce a small photon mass in the above expression to avoid infrared
divergences.Therefore we have
\begin{equation}
F=-e^{2}\int\frac{d^{3}ke^{ik\cdot x}}{(2\pi)^{2}}\theta(k^{0})[\delta
(k^{2}-\mu^{2})(\frac{m^{2}}{(r\cdot k)^{2}}+\frac{1}{(r\cdot k)}%
)-(d-1)\frac{\partial}{\partial\mu^{2}}\delta(k^{2}-\mu^{2})].
\end{equation}
If we use the parameter tric%
\begin{align}
\lim_{\epsilon\rightarrow0}\int_{0}^{\infty}d\alpha e^{-\alpha(\epsilon
-ik\cdot r)}  &  =\frac{i}{k\cdot r},\nonumber\\
\lim_{\epsilon\rightarrow0}\int_{0}^{\infty}\alpha d\alpha e^{-\alpha
(\epsilon-ik\cdot r)}  &  =-\frac{1}{(k\cdot r)^{2}},
\end{align}
and the retarded propagator with bare mass $\mu$%
\begin{align}
D^{+}(x)  &  =\int\frac{d^{3}k}{i(2\pi)^{2}}\delta(k^{2}-\mu^{2})\theta
(k^{0})e^{ik\cdot x}\nonumber\\
&  =\frac{1}{i(2\pi)^{2}}\int_{0}^{\infty}\frac{\pi kdkJ_{0}(k\left\vert
x\right\vert )}{2\sqrt{k^{2}+\mu^{2}}}=\frac{e^{-\mu|x|}}{8\pi i\left\vert
x\right\vert },
\end{align}
we obtain the function $F$ as
\begin{align}
F  &  =ie^{2}m^{2}\int_{0}^{\infty}\alpha d\alpha D_{+}(x+\alpha r)-e^{2}%
\int_{0}^{\infty}d\alpha D_{+}(x+\alpha r),\nonumber\\
&  =ie^{2}m^{2}F_{1}(x)-e^{2}F_{2}(x)
\end{align}
in the Feynman gauge.Soft photon divergence corresponds to the large $\alpha$
region and $\mu$ is an infrared cut-off$.$It is simple to evaluate the gauge
dependent term in $F$ by%
\begin{equation}
F_{L}=-\int\frac{d^{3}k}{(2\pi)^{2}}\theta(k_{0})\frac{\partial}{\partial
\mu^{2}}\delta(k^{2}-\mu^{2})e^{ik\cdot x}.
\end{equation}

Finally we get%
\begin{align}
F_{2}  &  =\int_{0}^{\infty}d\alpha\frac{\exp(-\mu(x+\alpha r))}{8\pi(x+\alpha
r)}=\frac{E_{1}(\mu|x|)}{8\pi\sqrt{r^{2}}},\\
F_{1}  &  =\int_{0}^{\infty}\alpha d\alpha\frac{\exp(-\mu(x+\alpha r))}%
{8\pi(x+\alpha r)}=\frac{\exp(-\mu|x|)-\mu|x|E_{1}(\mu|x|)}{8\pi r^{2}\mu},\\
F_{L}  &  =\frac{1}{16\pi}\frac{\exp(-\mu|x|)}{\mu},
\end{align}
where $r^{2}=m^{2}$ and
\begin{equation}
E_{1}(z)=\int_{z}^{\infty}\frac{\exp(-t)}{t}dt.
\end{equation}
Series expansion and asymptotoc expansion of $E_{1}(z)$ are%
\begin{align*}
E_{1}(z)  &  =-\gamma-\ln(z)-\sum_{n=1}^{\infty}\frac{(-1)^{n}z^{n}}%
{nn!},(|\arg z|<\pi)\\
E_{1}(z)  &  \sim\frac{\exp(-z)}{z}\{1-\frac{1}{z}+\frac{1\cdot2}{z^{2}}%
-\frac{1\cdot2\cdot3}{z^{3}}+...\},(|\arg z|<\frac{3}{2}\pi).
\end{align*}
From the above expressions,we have the short and long distance behaviour of
$E_{1}(\mu|x|)$
\begin{align*}
E_{1}(\mu\left\vert x\right\vert )  &  =-\gamma-\ln(\mu\left\vert x\right\vert
)+\mu|x|+O(\mu^{2})(\mu|x|\ll1),\\
E_{1}(\mu|x)  &  =\frac{\exp(-\mu|x|)}{\mu|x|}(1-\frac{1}{\mu|x|}+\frac
{2}{(\mu|x|)^{2}})(\mu|x|\gg1).
\end{align*}
For the leading order in $\mu$ we obtain%
\begin{align}
e^{2}m^{2}F_{1}  &  =\frac{e^{2}}{8\pi}(\frac{1}{\mu}+\left\vert x\right\vert
(-1+\ln(\mu\left\vert x\right\vert )+\gamma))+O(\mu^{3}),\\
-e^{2}F_{2}  &  =\frac{e^{2}}{8\pi m}(\ln(\mu\left\vert x\right\vert
)+\gamma-\mu|x|)+O(\mu),\\
(d-1)e^{2}F_{L}  &  =\frac{(d-1)e^{2}}{16\pi}(\frac{1}{\mu}-|x|)+O(\mu).
\end{align}
At short distance we have
\begin{align}
F  &  =\frac{(d+1)e^{2}}{16\pi\mu}-\frac{e^{2}}{8\pi m}\mu|x|+\frac{e^{2}%
}{8\pi m}(1+m|x|)(\ln(\mu\left\vert x\right\vert +\gamma)\nonumber\\
&  -\frac{(d+1)e^{2}}{16\pi}|x|.
\end{align}
Finally we define the function $F$ in Mikowski space by analytic continuation
from $\sqrt{-x^{2}}\rightarrow it$%
\begin{equation}
F=-\frac{e^{2}}{8\pi m}E_{1}(i\mu t)(itm+1)+\frac{(d+1)e^{2}}{16\pi\mu
}e^{-i\mu t}.
\end{equation}
If we expand in real time $t$,we obtain the short and long time behaviour%
\begin{equation}
F=\frac{e^{2}}{8\pi m}(\gamma+\ln(i\mu t))+\frac{(d+1)e^{2}}{16\pi\mu
}+O(t)),|\mu t|\ll1,
\end{equation}%
\begin{equation}
F\simeq\frac{e^{2}}{8\pi m}(\frac{me^{-i\mu t}}{\mu}-\frac{e^{-i\mu t}}{i\mu
t})+\frac{(d+1)}{16\pi\mu}e^{-i\mu t},1\ll|\mu t|,
\end{equation}
provided%
\begin{equation}
E_{1}(i\mu t)=\frac{-ie^{-i\mu t}}{\mu t}+O\text{ }(\frac{1}{t^{2}}),|\mu
t|\ll1.
\end{equation}

\subsection{Feynman-Dyson perturbative spectral function}

In the Feynman-Dyson theory $O(e^{2})$ propagator is given[5]
\begin{equation}
S_{F}^{^{\prime}}(p)=S_{F}(p)-S_{F}(p)\sum(p)S_{F}(p),
\end{equation}
where
\begin{equation}
\sum(p)=-\frac{e^{2}}{16\pi}\int_{-\infty}^{\infty}\frac{da}{\gamma\cdot
p-a+i\epsilon}[\frac{4m}{a}+d\frac{(a+m)^{2}}{a^{2}}]\theta(a^{2}-m^{2}).
\end{equation}
From the above formula we obtain the self-energy%
\begin{equation}
\sum(p)=\frac{e^{2}}{8\pi}[\frac{m}{\sqrt{p^{2}}}(2+d)\ln(\frac{m-\sqrt{p^{2}%
}}{m+\sqrt{p^{2}}})-d\gamma\cdot p(\frac{1}{2\sqrt{p^{2}}}(1+\frac{m^{2}%
}{p^{2}})\ln(\frac{m-\sqrt{p^{2}}}{m+\sqrt{p^{2}}})+\frac{1}{m})].
\end{equation}%
\begin{align}
-S_{F}(p)\sum(p)S_{F}(p)  &  =\frac{de^{2}}{8\pi}\frac{\gamma\cdot p+m}%
{p^{2}-m^{2}}(\frac{1}{2\sqrt{p^{2}}}(1+\frac{m^{2}}{p^{2}})\ln(\frac
{m-\sqrt{p^{2}}}{m+\sqrt{p^{2}}})+\frac{1}{m})\nonumber\\
&  +\frac{p^{2}+3m^{2}+2m\gamma\cdot p}{(p^{2}-m^{2})^{2}}\frac{e^{2}m}{8\pi
}(\frac{2+d}{\sqrt{p^{2}}}+\frac{d}{2\sqrt{p^{2}}})\ln(\frac{m-\sqrt{p^{2}}%
}{m+\sqrt{p^{2}}})+d).
\end{align}
In this approximation $O(e^{2})$ spectral functions for vector and scalar part
are given%
\begin{equation}
S_{F}(p)\sum(p)S_{F}(p)=\int_{(m+\mu)^{2}}^{\infty}\frac{ds(\rho_{1}%
(s)\gamma\cdot p+\rho_{2}(s))}{p^{2}-s+i\epsilon}.
\end{equation}%
\begin{align}
\rho_{1}(s)  &  =-\frac{e^{2}}{8\pi}(\frac{2m^{2}}{(s-m^{2})^{2}}\frac
{4+3d}{2\sqrt{s}}+\frac{d}{(s-m^{2})2\sqrt{s}}(1+\frac{m^{2}}{s}%
))\theta(s-m^{2}),\\
\rho_{2}(s)  &  =-\frac{e^{2}}{8\pi}(\frac{(s+3m^{2})m}{(s-m^{2})^{2}}%
\frac{4+3d}{2}+\frac{md}{(s-m^{2})2}(1+\frac{m^{2}}{s}))\theta(s-m^{2}).
\end{align}
It is now clear that the above spectral functions have different gauge
dependences from that ones obtaind based on the definition (73) in section II
B 3.In 4-dimension it is known that the Feynman-Dyson theory leads the same
spectral function as we get based on its definition.Formal proof is given for
simple scalar model in ref[5].

\end{document}